\documentclass[twocolumn,aps,prl,preprintnumbers,floatfix]{revtex4}
\usepackage{bbm}
\usepackage{bm}
\usepackage{amsmath}
\usepackage{amssymb}
\usepackage{empheq}
\usepackage{graphicx}
\usepackage{mathrsfs}
\usepackage{amsfonts}
\usepackage{amsthm}
\usepackage{color}
\usepackage{bigints}
\usepackage{txfonts}
\usepackage{hyperref}
\hypersetup{
     unicode=false,          % non-Latin characters in Acrobat?s bookmarks
     pdftoolbar=true,        % show Acrobat?s toolbar?
     pdfmenubar=true,        % show Acrobat?s menu?
     pdffitwindow=false,     % window fit to page when opened
     pdfstartview={FitH},    % fits the width of the page to the window
     pdftitle={My title},    % title
     pdfauthor={Author},     % author
     pdfsubject={Subject},   % subject of the document
     pdfcreator={Creator},   % creator of the document
     pdfproducer={Producer}, % producer of the document
     pdfkeywords={keyword1} {key2} {key3}, % list of keywords
     pdfnewwindow=true,      % links in new PDF window
     colorlinks=false,       % false: boxed links; true: colored links
     linkcolor=red,          % color of internal links (change box color with linkbordercolor)
     citecolor=green,        % color of links to bibliography
     filecolor=magenta,      % color of file links
     urlcolor=cyan           % color of external links
}

\providecommand{\U}[1]{\protect\rule{.1in}{.1in}}
\makeatletter
\@ifundefined{textcolor}{}
{
\definecolor{BLACK}{gray}{0}
\definecolor{WHITE}{gray}{1}
\definecolor{RED}{rgb}{1,0,0}
\definecolor{GREEN}{rgb}{0,1,0}
\definecolor{BLUE}{rgb}{0,0,1}
\definecolor{CYAN}{cmyk}{1,0,0,0}
\definecolor{MAGENTA}{cmyk}{0,1,0,0}
\definecolor{YELLOW}{cmyk}{0,0,1,0}
}
\makeatother

\begin{document}

\title{Non-Wigner-Dyson level statistics and fractal wavefunction of disordered Weyl semimetals}
\author{C. Wang$^{1,2}$, Peng Yan$^{1}$, and X. R. Wang $^{2,3}$}
\email[Corresponding author: ]{phxwan@ust.hk}
\affiliation{$^{1}$School of Electronic Science and Engineering
and State Key Laboratory of Electronic Thin Film and
Integrated Devices, University of Electronic Science and Technology
of China, Chengdu 610054, China}
\affiliation{$^{2}$Physics Department, The Hong Kong University of Science and Technology,
Clear Water Bay, Kowloon, Hong Kong}
\affiliation{$^{3}$HKUST Shenzhen Research Institute, Shenzhen 518057, China}
\date{\today}

\begin{abstract}
Finding fingerprints of disordered Weyl semimetals (WSMs) is an unsolved task. 
Here we report such findings in the level statistics and the fractal nature of 
electron wavefunction around Weyl nodes of disordered WSMs. 
The nearest-neighbor level spacing follows a new universal distribution $P_c(s) 
=C_1 s^2\exp[-C_2 s^{2-\gamma_0}]$ originally proposed for the level statistics 
of critical states in the integer quantum Hall systems or normal dirty metals 
(diffusive metals) at metal-to-insulator transitions, instead of the Wigner-Dyson 
distribution for diffusive metals. Numerically, we find $\gamma_0=0.62\pm0.07$. 
In contrast to the Bloch wavefuntions of clean WSMs that uniformly distribute over 
the whole space of ($D=3$) at large length scale, the wavefunction of disordered 
WSMs at a Weyl node occupies a fractal space of dimension $D=2.18\pm 0.05$. 
The finite size scaling of the inverse participation ratio suggests 
that the correlation length of wavefunctions at Weyl nodes ($E=0$) diverges as 
$\xi\propto |E|^{-\nu}$ with $\nu=0.89\pm0.05$. In the ergodic limit, the level 
number variance $\Sigma_2$ around Weyl nodes increases linearly with the average 
level number $N$, $\Sigma_2=\chi N$, where $\chi= 0.2\pm0.1$ is independent 
of system sizes and disorder strengths. 
\end{abstract}

\maketitle

Crystal Weyl semimetals (WSMs), characterized by the linear crossings of 
their conduction and valence bands at Weyl nodes (WNs) and topologically 
protected surface states, have attracted enormous attention in recent 
years because of their exotic properties and potentials in applications 
\cite{wanxg,yangky,burkovaa,wenghm,xusy,Lul,shekhar,burkov}. 
There is little doubt that WSMs are a new state of matter in nature.  
However, the characteristics of WSMs for a crystal do not apply to a 
disordered system since the lattice momentum is not a good quantum number.  
To fully establish the WSMs as a genuine state of matter, one needs to 
find their fingerprints in disorders that inevitably exist in all materials. 
It was originally believed that disordered WSMs are featured by vanishing 
density of states (DOS) at WNs \cite{Q5,Q6,Q7,royb,beras}. 
The divergence of the bulk state localization length at the Weyl-semimetal-to-diffusive-metal
(WSM-to-DM) transition was also conjectured \cite{Q1,Q2,Q3,chencz1,shapourianh}. 
Nonetheless, these features were challenged in many recent studies 
\cite{Q4,pixleyjh2,suy,Ziegler1}. So far, a simple working criterion 
for disordered WSMs is still lacking. This study aims to search the 
fingerprints from the random matrices that describe the disordered WSMs. 

Random matrices have broad applications in many fields of physics 
\cite{wingerep,verbaarschotjj,ffranchini,sanchezd,bohigaso}. 
%the nuclear physics \cite{wingerep}, the quantum chromodynamics \cite{verbaarschotjj}, 
%the quantum gravity \cite{ffranchini}, the condensed matter physics \cite{janssenm,sanchezd}, 
%the quantum optics \cite{aaronsons}, and the quantum chaos \cite{bohigaso}.
In the condensed matter physics, random matrix theory can be used 
to distinguish different types of metals. % from the Anderson insulators. 
The distribution $P(s)$ of nearest-neighbour level spacing $s$ (in the unit of 
the mean level spacing) of diffusive metals (normal dirty metals) and the level 
number variance $\Sigma_2(\Delta E)=\langle n^2\rangle-\langle n\rangle^2$ in 
a given energy range are governed by the Wigner-Dysion distributions \cite{mehta}. 
For example, $P(s)$ follows the Wigner surmises, $P_{\beta}(s)= C_1s^{\beta}\exp
[-C_2s^2]$ ($C_1$ and $C_2$ are determined by the probability normalization and 
the unity of mean level spacing $\int P_{\beta}(s) ds=\int sP_{\beta}(s)ds= 1$), 
where $\beta=1$, 2 and 4, depending on symmetries \citep{symmetry}. 
%$\beta=1$ for systems with time-reversal (TR) symmetry and spin-rotation symmetry is 
%called the Gaussian orthogonal ensemble; $\beta=2$ for systems without TR symmetry 
%is called the Gaussian ensemble (GUE); $\beta=4$ for systems with TR 
%symmetry and no spin-rotation symmetry is called the Gaussian symplectic ensemble. 
They are in contrast to the Poisson distribution $P_{\text{Loc}}(s)=\exp[-s]$ for 
level spacings of Anderson insulators. One interesting and fundamental question 
is whether level statistics can be used to distinguish disordered WSMs from  
normal dirty metals. 
%what kind of level statistics of the newly discovered disordered Weyl semimetals 
%(WSMs) must follow, especially around the Weyl nodes (WNs).  
\par

Far from the WNs, the level statistics of extended states of a disordered 
WSM should not behave differently from those of diffusive metals (DMs).
However, near the WNs, the numbers of extended states are few \cite{dos}. 
In another word, extended states tend to avoid these points besides their general 
level repulsion effect. Thus, this ``double'' repulsion should result in weak decay 
of $P(s)$ in the tail. On this aspect, WNs are very similar to the critical points 
located near each Landau subband \cite{xiongg,wangc} of the integer quantum Hall (IQH) 
systems where there is only one extended state, or the conventional metal-insulator 
transition point where the density of extended states is also vanishingly small. 
The level statistics of the extended states in IQH systems and near a metal-insulator 
transition point are believed to follow a new {\it critical} level statistics 
$P_c(s)$ \cite{altshulerb1,shklovskiibi,kravtsovve,aronovga,admirlin,Garcia,klesseb}.
For Gaussian unitary ensemble (GUE) \cite{symmetry}, a sub-Gaussian decay of $P_c(s)$ and 
a linear increase of $\Sigma_2(\Delta E)$ with the mean level number $N$ were predicted:
\begin{equation}
\begin{gathered}
P_c(s)=C_1 s^2\exp[-C_2 s^{2-\gamma_0}],
\end{gathered}\label{novel_ls_2}
\end{equation}
and
 \begin{equation}
\begin{gathered}
\Sigma_2(\Delta E)=\chi N,
\end{gathered}\label{novel_ls_3}
\end{equation}
that differ from the Wigner-Dyson distributions mentioned early.
Here $\gamma_0=1-1/(\nu d)$ with the correlation length critical exponent 
$\nu$ and the spatial dimension $d$. $\chi>0$ is a universal constant, known 
as the ``spectral compressibility''. Naturally, one suspects that extended 
states in the vicinity of the WNs follow the critical level statistics. 
The focus of the current work is to see whether this conjecture is 
correct or not, and whether it can be a fingerprint of disordered WSMs.
\par

In this work, we use the exact diagonalization method to compute the eigenenergies 
and eigenfunctions of a disordered WSM on a cubic lattice. Through the 
finite-size scaling analysis of the inverse participation ratio (IPR), we find 
that the wavefunctions at WNs ($E=0$) are fractals of dimension $2.18\pm0.05$. 
The finite size scaling of the IPR reveals the correlation length $\xi(E)$ of 
wavefunctions around WNs diverging as $\xi=\xi_0(W)|E/t|^{-\nu}a$ with 
$\nu=0.89\pm0.05$, where $a$ and $t$ are respectively the lattice constant and 
the hopping energy. $\xi_0(W)$ is a disorder-dependent dimensionless coefficient.
For a finite system of size $L$, the energy level spacing distribution $P(s)$ 
within the energy window of $|E|<\epsilon_c=t(\xi_0 a/L)^{1/\nu}$ follows the 
critical distribution $P_c(s)$, while the level spacing of extended states outside 
the energy window is described by the Wigner-Dyson distribution $P_{\beta=2}(s)$. 
For a small energy range $\Delta E$ within which the average level number $N$ is 
small, the level number variance $\Sigma_2(\Delta E)$ around WNs is linear in $N$, 
$\Sigma_2(\Delta E)=\chi N$, with a disorder-independent constant $\chi=0.2\pm0.1$. 
At moderate disorders where the pair of WNs annihilate each other, $\Sigma_2(\Delta E)$
increases logarithmically with $N$, in agreement with the Wigner-Dyson prediction
for GUE.  
\par 

\emph{Model and methods.$-$}To substantiate our claims, we consider a tight-binding 
Hamiltonian on a cubic lattice of size $L^3$ and unity lattice constant $a=1$, 
\begin{equation}
\begin{gathered}
H=H_0+V.
\end{gathered}\label{model_1}
\end{equation}
The first term is for a pure system,  
\begin{equation}
\begin{gathered}
H_0=\sum_i m_z c^\dagger_i\sigma_z c_i-\sum_i \left[
\dfrac{m_0}{2}\left(c^\dagger_{i+\hat{x}}\sigma_z c_i+
c^\dagger_{i+\hat{y}}\sigma_z c_i\right)\right.\\
+\left.\dfrac{t}{2}\left(c^\dagger_{i+\hat{z}}\sigma_z c_i
+ic^\dagger_{i+\hat{x}}\sigma_x c_i
+ic^\dagger_{i+\hat{y}}\sigma_y c_i+H.c.\right)\right],
\end{gathered}\label{model_2}
\end{equation}
where $c^{\dagger}_i=(c^{\dagger}_{i\uparrow},c^{\dagger}_{i\downarrow})$ and 
$c_i$ are the single electron creation and annihilation operators at site $i$. 
$\hat{x},\hat{y},\hat{z}$ are the unit lattice vectors along the $x,y,z$ 
directions, respectively. $\sigma_{x,y,z}$ are the Pauli matrices for spin. 
$m_z$ and $m_0$ are respectively Dirac and Newtonian masses. 
%The Hamiltonian \eqref{model_2} can be blocked diagonalized in the momentum space, 
%$H_0=\sum_{\bm{k}}c^{\dagger}_{\bm{k}}\mathcal{H}_{\bm{k}}c_{\bm{k}}$, where 
%$\mathcal{H}_{\bm{k}}=(m_z-t\cos k_z-m_0\cos k_x-m_0\cos k_y)\sigma_z+t(\sin 
%k_x\sigma_x+\sin k_y\sigma_y)$. 
Hereafter we take $m_0/t=2.1$ and $m_z/t=0$ with $t$ being the energy unit 
such that $H_0$ is a WSM with two pairs of WNs at $E=0$ \cite{chencz1}.
Disorders are introduced through the second term
\begin{equation}
\begin{gathered}
V=\sum_i V^0_i\sigma_0+V^z_i\sigma_z,
\end{gathered}\label{model_3}
\end{equation}
where $V^0_i$ and $V^z_i$ distribute randomly and uniformly in the range of 
$[-W/2,W/2]$ such that $W$ measures the randomness strength. Of course, one could 
also introduce the disorders by other terms such as $V=\sum_i V^x_i\sigma_x+
V^y_i\sigma_y$, but the physics does not change \cite{supp}. 
% As shown in the Supplemental Materials \cite{supp}, the 
% physics does not change.
\par 

Hamiltonians \eqref{model_2} and \eqref{model_3} were used in 
Refs.~\cite{chencz1,suy,beras,Q2,shapourianh} to study phase 
transitions from disordered WSMs to various other quantum phases. 
Disordered WSMs in this model break the time-reversal symmetry and 
belong to the GUE with symmetric index $\beta=2$ \cite{symmetry}.
% According to the usual classification, this model belongs to the GUE. 
Early numerical calculations \cite{Q3,chencz1,suy} of localization lengths 
and averaged Hall conductivities have established following results for 
the model when the Fermi level is at $E=0$: 
1) In the weak disorder of $W\leq W_{c,1}\simeq 5.2$, the system is a disordered WSM.
2) For the intermediate disorder of $W_{c,3}\simeq 21 >W>W_{c,2}\simeq 6.2$,
the system is a normal (without topological surface states) DM.
3) For strong disorder of $W>W_{c,3}$, the state of $E=0$ is localized, and the
system is an Anderson insulator. Whether there is a phase transition between 
$W_{c,1}$ and $W_{c,2}$ is an issue under debate \cite{suy}. 
Later, we shall provide convincing evidence in the level statistics that 
this is a direct WSM-to-DM transition.
\par

We use the exact diagonalization method to obtain all eigenenergies and eigenfunctions.
The IPR, defined as $p_2(E)=\langle\sum_i|\psi_i(E)|^4\rangle^{-1}$ where $\psi_i(E)$ 
is the amplitude of normalized wavefunction of energy $E$ at site $i$, is used to study 
the wavefunction structure. IPR measures the number of sites that the state occupies
and scales with sizes as $p_2\sim\text{const}$ [$p_2\propto L^d$] for a localized 
(normal extended) state \cite{xrwang}.
%and scales with $L$ as $p_2\propto L^{D}$ for a fractal wavefunction of dimension $D$ 
%while $p_2\sim\text{const}$ [$p_2\propto L^d$] for a localized (normal extended) state 
%\cite{xrwang}.
If there exists an isolated critical state at $E_c$ whose wavefunction is a fractal, the 
one-parameter scaling analysis says that IPR of the states near $E_c$ is a universal 
function of the fractal dimension $D$ ($0<D<d$) \cite{Q7,xrwang,janssenm},
\begin{equation}
\begin{gathered}
p_2(E)=L^{D}[f(L/\xi)+C_3/L^y],
\end{gathered}\label{pr}
\end{equation}
where $f(x)$ is an unknown scaling function, $C_3$ is the coefficient of the 
finite-size correction, and $y>0$ is the exponent for the irrelevant variable. 
The correlation length $\xi$ diverges as $\xi=\xi_0|E-E_c|^{-\nu}$, 
where $\nu$ is the critical exponent characterizing the universality class.
The state wavefunction occupies a fractal (whole) space of dimensionality $D\neq 3$ 
($d=3$) if the system size is smaller (larger) than $\xi$. 
\par

In our analysis, we fit the numerically obtained $p_2(E)$ by Eq.~\eqref{pr} \cite{supp}. 
The identification of whether a state is extended, localized, or critical is 
guided by the following criteria: 
(1) For extended (localized) states, $Y_L(E)=p_2 L^{-D}-C_3/L^y$ increases 
(decreases) with system size $L$; (2) For critical states, $Y_L(E)=f(L/\xi)$ is 
the one-parameter scaling function and size-independent for $E$ close to $E_c$. 
\par

%\emph{Nearest-level spacing distribution $P(s)$.$-$}
To compute the level statistics of states around the WNs, we diagonalize 
the Hamiltonian \eqref{model_1} by imposing periodic boundary conditions in
all directions in order to eliminate the edge state effects. 
% We consider the eigenenergies $\{ E_j\}$ in a very narrow energy window for many realiztions. 
Near the WNs, the DOSs decrease with $|E|$ algebraically such that 
a proper renormalization is needed to correctly compute $P(s)$. 
We also eliminate the systematic error in the histogram plots to increase the 
accuracy of $P(s)$ \cite{wangc2}.
\par 

\begin{figure}[ht!]
\centering
  \includegraphics[width=0.45\textwidth]{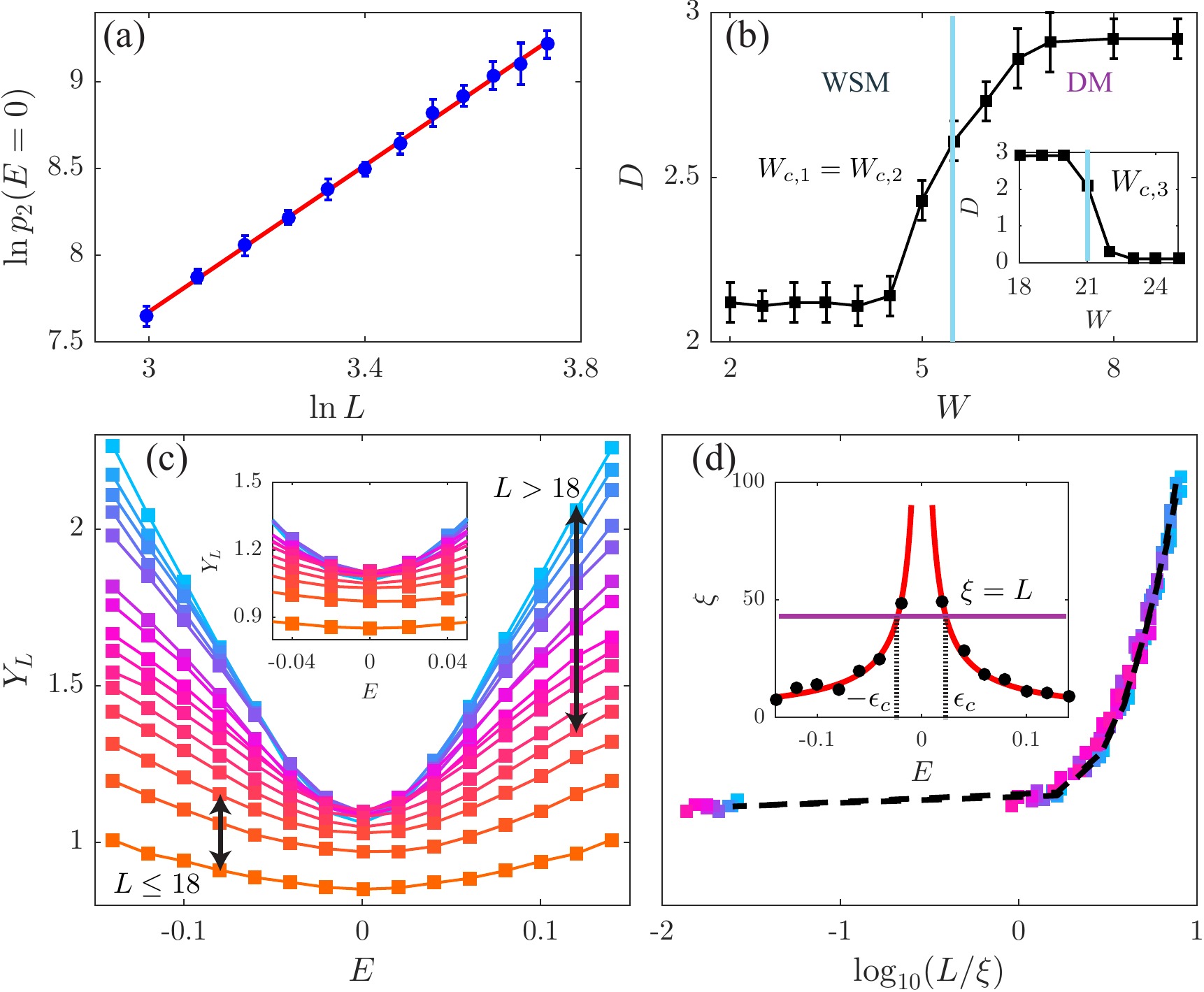}
\caption{(a) $\ln p_2(E=0)$ v.s. $\ln L$ for $W=4$.  
The solid line is the linear fit of slope $D=2.18\pm0.05$. (b) $D$ v.s. $W$ for $2\leq 
W\leq 9$ and $18\leq W\leq 24$ (Inset). (c) $Y_L$ v.s. $E$ for $L=42,40,38,\cdots,14$ 
(from up to down) for $W=4$. (d) Scaling function $f(x=\log_{10}(L/\xi))$.
Black dash line connects points to guide eyes. Inset: $\xi(E,W=4)$ v.s. $E$. 
Red solid line is the fit of $\xi=\xi_0|E|^{-\nu}$. 
Dash lines locate $-\epsilon_c(L)$ and $\epsilon_c(L)$. 
Error bars for all the data points are smaller than symbol sizes in (c) and (d).}
\label{fig_fd}
\end{figure}

\emph{Results of IPR.$-$}Figure~\ref{fig_fd}(a) 
displays $\ln p_2(E=0)$ v.s. $\ln L$ at $W=4<W_{c,1}$. The curve is 
a straight line with a slope (fractal dimension) of $D=2.18\pm0.05$, 
in contrast to a normal extended state that occupies the whole space. 
Wavefunctions at WNs have a universal fractal structure in the sense 
that $D$ does not depend on the disorder strength $W<W_{c,1}$. 
This is clearly shown by the $D-W$ curve in Fig.~\ref{fig_fd}(b). 
It is evident that $D\simeq 2.18$ is constant for $W<W_{c,1}$. 
However, for $W_{c,3}>W>W_{c,2}$, we find $D\simeq 3$ that indicates the state 
of $E=0$ becomes a normal extended state, and the system becomes a DM. 
Above $W_{c,3}$, the zero energy state is localized with $D\simeq 0$, 
see the inset of Fig.~\ref{fig_fd}(b). Our calculations of $D$ are 
consistent with the direct WSM-to-DM transition. 
\par

To confirm the criticality of WNs, we perform the chi-square fit of 
$p_2(E)$ to Eq.~\eqref{pr} and plot $Y_L(E)$ for $W=4$ in Fig.~\ref{fig_fd}(c). 
If only systems of small sizes are considered, say $L\leq 18$, $Y_L(E)$ always 
increases with $L$, and one would conclude that all states are extended. 
However, we find $dY_L/dL=0$ at WNs for $L>18$ within numerical errors, 
instead of $dY_L/dL>0$ for an extended state. 
The merging of $Y_L(E=0)$ indicates $E=0$ is a critical point and a novel phase 
transition between an isolated critical state at WNs and extended states 
at $E\neq 0$, very similar to the phase transition from an isolated critical 
level to localized states in IQH systems \cite{wangc}.
\par

Our data fit well to the one-parameter scaling hypothesis of 
$Y_L(E)=f(L/\xi)$ with $\xi=\xi_0|E|^{-\nu}$ for states near $E=0$. 
From the chi-square fit of $p_2(E)$ for $L>18$, we obtain $\nu=0.89\pm 
0.05$, $y=0.4\pm0.1$, $\xi_0=1.5\pm0.3$, and $C_3=3.2\pm0.1$ \cite{supp}. 
The goodness-of-fit $Q=0.3$ is a quite satisfactory number, thus it supports 
$E=0$ as a quantum critical point separating a critical state from 
extended states. The smooth scaling function $f(x=\log_{10}(L/\xi))$ 
with $\xi=\xi_0|E|^{-\nu}$ is shown in Fig.~\ref{fig_fd}(d) obtained 
by collapsing all $Y_L(E)$ curves of different $L$ into a single curve.
\par

\begin{figure}[ht!]
\centering
  \includegraphics[width=0.43\textwidth]{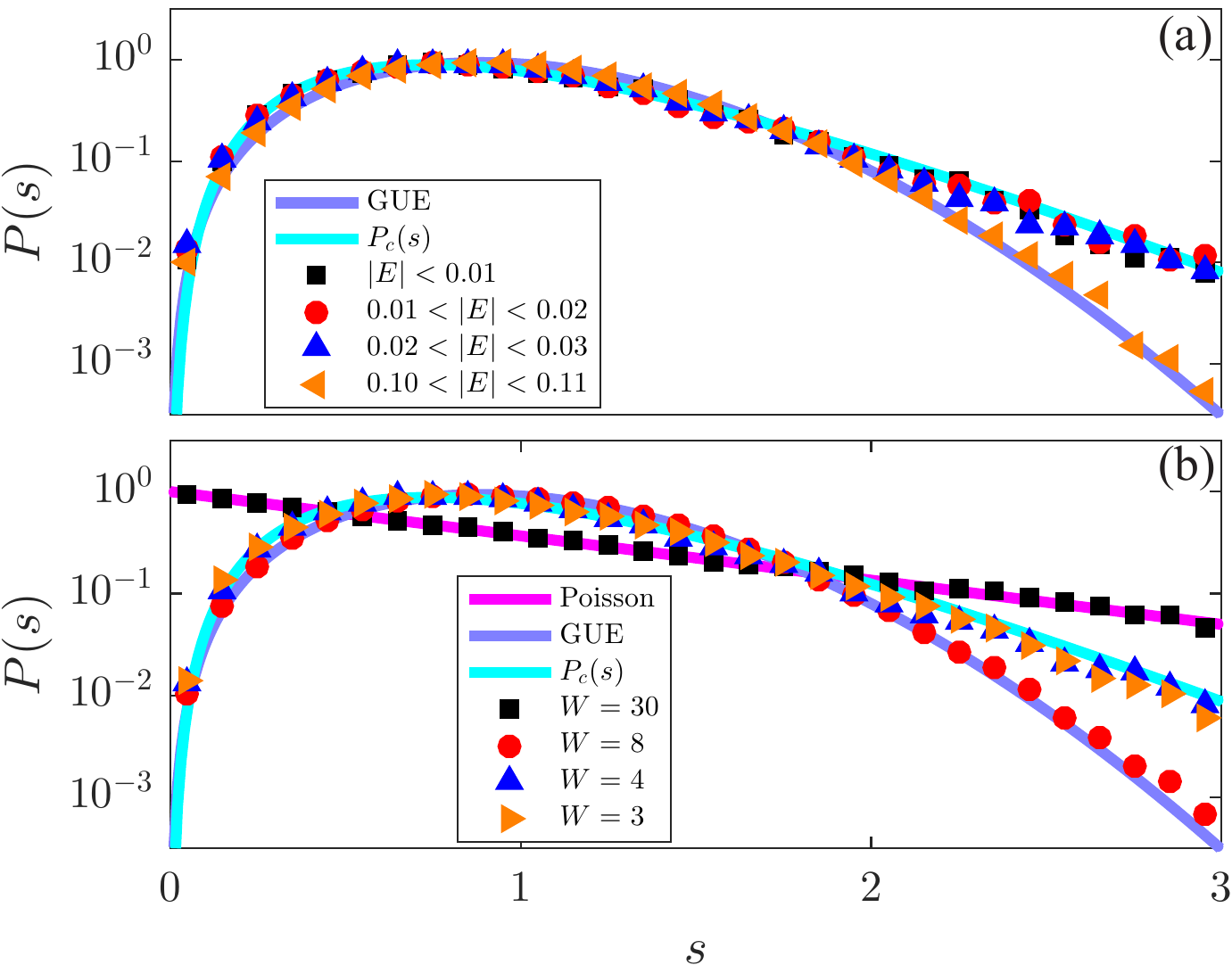}
\caption{(a) $P(s)$ for different energy windows at $W=4$. 
The cyan and slate-blue lines are Eq.~\eqref{novel_ls_2} and $P_{\beta=2}(s)$, respectively. 
(b) $P(s)$ for $W=3,4,8,30$ in a fixed energy window $E/t\in[-0.03,0.03]$:
For $W=3,4<W_{c,1}$, numerical data of $P(s)$ agrees with $P_c(s)$. 
For $W_{c,3}>W=8>W_{c,2}$, $P(s)$ data falls on $P_{\beta=2}(s)$. 
For $W=30>W_{c,3}$, $P(s)$ accords with $P_{\text{Loc}}(s)$. Here $L=30$.}
\label{fig_ps}
\end{figure}

\emph{Results of $P(s)$.$-$}Because $E=0$ is an isolated critical point 
for $W<W_{c,1}$, one should expect that all states for a system of size $L$ 
within a small energy range of $|E|<\epsilon_c$ indicated by the vertical dash 
lines in the inset of Fig.~\ref{fig_fd}(d), where the critical energy is defined 
as $\epsilon_c=(\xi_0/L)^{1/\nu}$, look like fractals of dimension $D=2.18$. 
The level spacing distribution $P(s)$ within the energy window is expected to follow 
the critical level statistics of Eq.~\eqref{novel_ls_2} coming from the assumption 
of power-law decay of wavefunctions \cite{altshulerb1,kravtsovve,aronovga,admirlin}.
Our conjecture is confirmed by Fig.~\ref{fig_ps}(a). Apparently, $P(s)=P_c(s)$ 
only for states near WNs ($|E|<\epsilon_c=(\xi_0(W=4)/L)^{1/\nu}=0.035$), while 
far away from the WNs (say $|E|\in [0.10,0.11]$), $P(s)=P_{\beta=2}(s)$. 
We note also that $P(s)$ evolves from $P_{\beta=2}(s)$ to $P_c(s)$ 
as the energy window approached $[-\epsilon_c,\epsilon_c]$. 
Thus, although $P_c(s)$ persists at $|E|<\epsilon_c$ for finite $L$, it will 
happen only at WNs in the thermodynamics limit $L\rightarrow\infty$.
Similar features have been observed by a fixed energy window and varying $L$ 
($\epsilon_c$) \cite{supp}.
\par

We investigate now how $P(s)$ within energy range of $[-0.03,0.03]$ changes with 
randomness strength $W$ from the universal level distribution $P_c(s)$ at a weak 
disorder, where zero energy wavefunction is a fractal, to the Poisson distribution 
at extremely strong disorder, where the zero energy wavefunction is localized. 
Some representative results are shown in Fig.~\ref{fig_ps}(b).
At an extremely strong disorder $W=30>W_{c,3}$ where all states are localized, 
$P(s)$ (black squares) follows the Poisson statistics $P_{\text{Loc}}$ 
(magenta line). While for $W_{c,3}>W=8>W_{c,2}$ such that the system is in the DM phase, 
$P(s)$ (red circles) follows the Wigner surmises $P_{\beta=2}(s)$. 
When the systems are WSMs, say $W=4,3 <W_{c,1}$, $P(s)$ are described 
by $P_c(s)$. 
%For $W\in [W_{c,1},W_{c,2}]$, we observe some intermediate level distribution between 
% $P_{\beta=2}(s)$ and $P_{c}(s)$ \cite{supp}. 
\par

\begin{figure}[ht!]
\centering
  \includegraphics[width=0.43\textwidth]{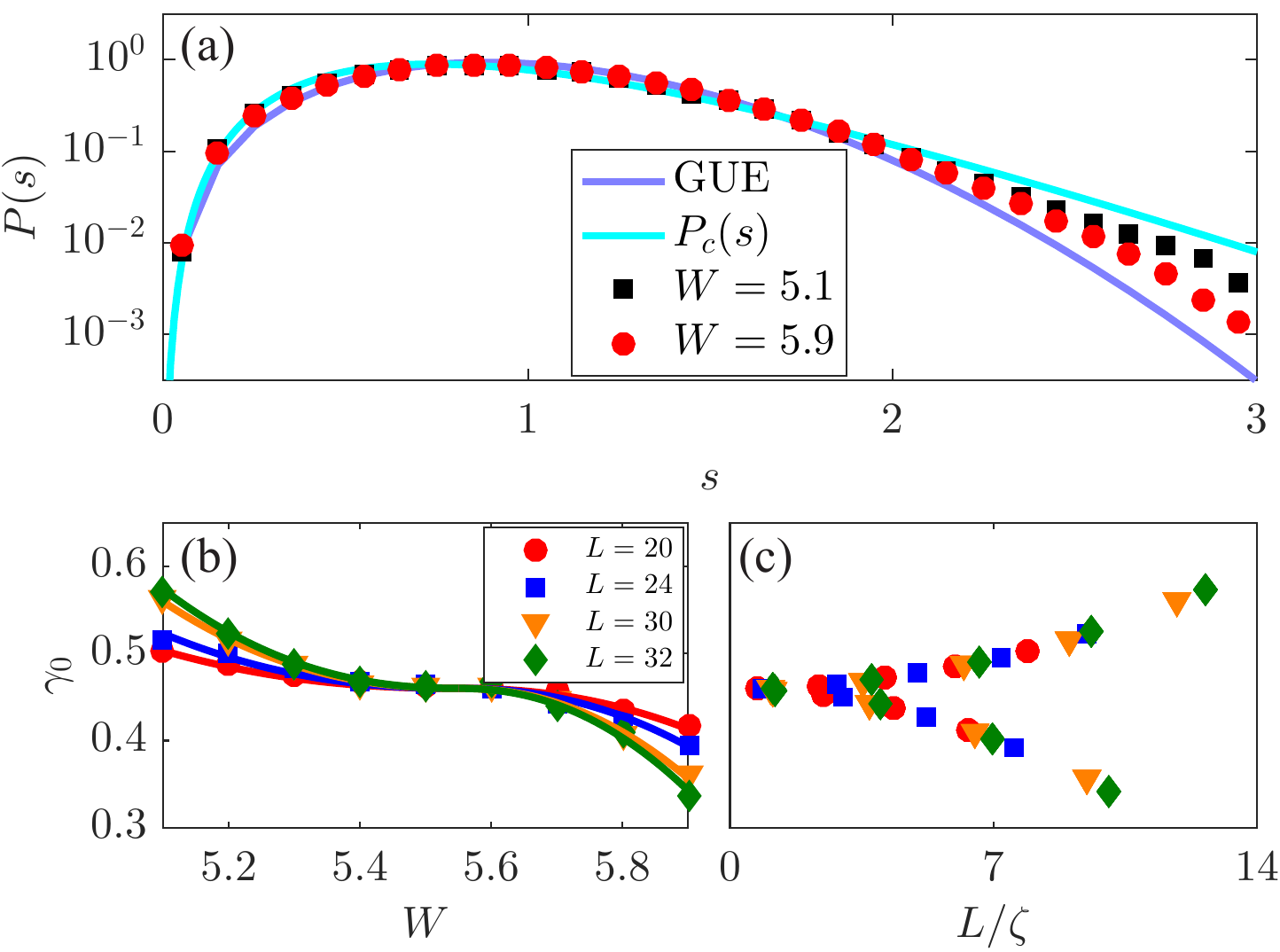}
\caption{(a) $P(s)$ for $W=5.1$ and 5.9 at $L=30$ and $|E|<0.03$. 
Cyan and slate-blue lines are the same as those in Fig.~\ref{fig_ps}. 
(b) $\gamma_0$ v.s. $W$ for $L=20,24,30,32$. (c) Scaling function $g(x)$ 
of $\gamma_0(W,L)$. 
}
\label{fig_scaling}
\end{figure}

Let us use above results to address the issue whether there is a direct 
WSM-to-DM transition in the regime of $W_{c,1}<W<W_{c,2}$. If there is a topological 
insulator phase between disordered WSM and DM phases \cite{suy}, one should expect 
$P(s)$ passing through the Poisson distribution on its way from $P_{c}(s)$ 
to $P_{\beta=2}(s)$. Figure~\ref{fig_scaling}(a) are two examples that 
show clearly $P(s)$ changing from $P_{c}(s)$ to $P_{\beta=2}(s)$ in 
the regime of $W_{c,1}<W<W_{c,2}$ without any sign of Poisson distribution. 
To further substantiate the assertion, we evaluate $\gamma_0(W,L)$ 
and show the results in Fig.~\ref{fig_scaling}(b). 
The transitions between the new universal level statistics ($\gamma_0=0.62$)  
and the Winger-Dyson distribution ($\gamma_0=0$) become sharper for larger $L$.
Besides, $\gamma_0(W,L)$ curves for different $L$ cross at a single disorder
$W=W_c\simeq 5.5$, indicate a quantum phase transition. We thus identify $W_c$ 
as the transition point that is substantiated by the nice collapse of all data 
in the vicinity of $W_c$ into a single parameter scaling function of  
\begin{equation}
\begin{gathered}
\gamma_0(W,L)=g(L/\zeta)
\end{gathered}\label{scaling}
\end{equation}
with the correlation length diverging as $\zeta=\zeta_0|W-W_c|^{-\nu}$, 
as shown in Fig. \ref{fig_scaling}(c). Obviously, two different branches 
for $W<W_c$ and $W>W_c$ correspond to two distinct phases. 
We thus consider Fig. \ref{fig_scaling} as an empirical verification 
of the existence of WSM-to-DM quantum phase transitions such that states of $W<W_c$ ($W>W_c$) 
belong to the WSMs (the DMs) for $L\to\infty$. 

\begin{figure}[ht!]
\centering
  \includegraphics[width=0.43\textwidth]{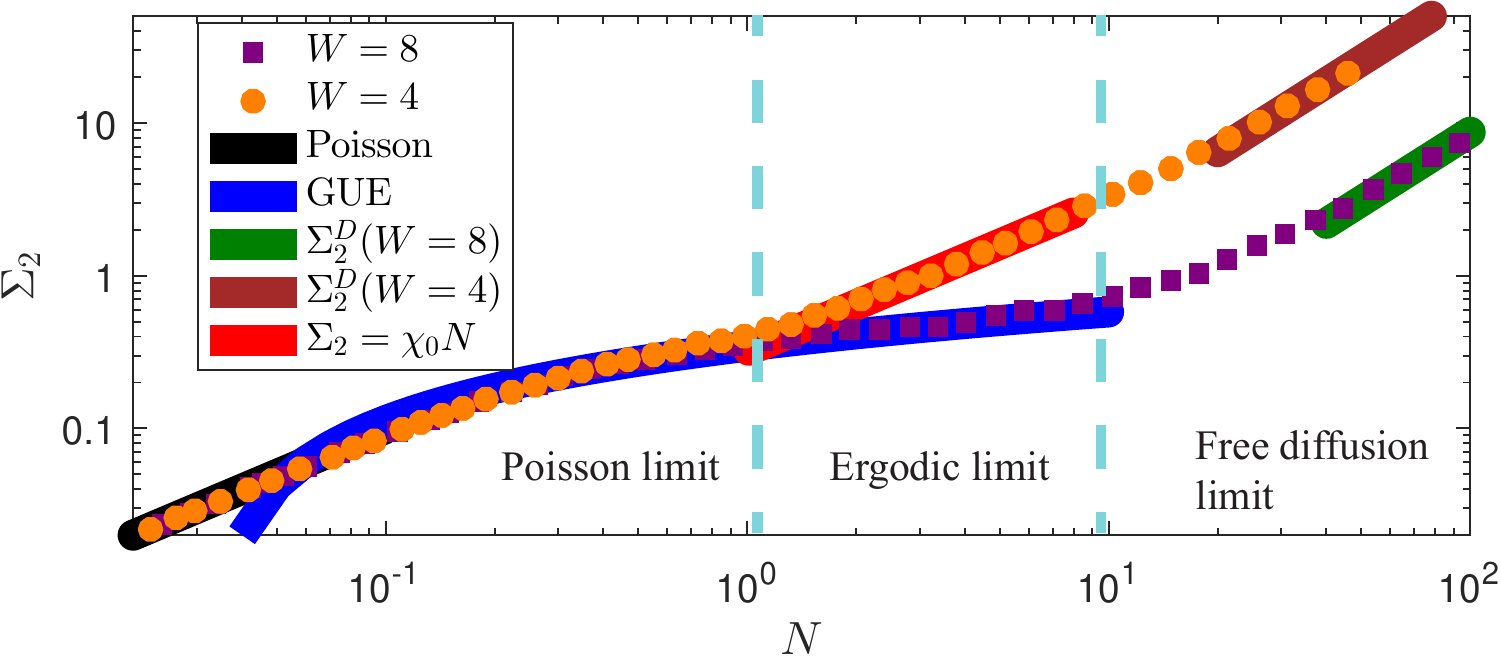}
\caption{$\Sigma_2$ v.s. $ N$ for $W=4$ (orange circles) and 8 (purple squares). 
Here $L=30$.
  }
\label{fig_variance}
\end{figure}  

\emph{Results of $\Sigma_2(\Delta E)$.$-$}We also compute the number variance $\Sigma_2
(\Delta E)$ for various energy ranges $[-\Delta E/2,\Delta E/2]$ (around WNs) and disorders.
Figure~\ref{fig_variance} displays $\Sigma_2(\Delta E)$ vs $N$ for WSM ($W=4$) 
and DM ($W=8$). According to the orthodox theory \cite{admirlin,altshulerb1,book}, 
there exist two important energy scales: the Thouless energy $E_T$ related to the 
dimensionless conductance of a mesoscopic system and the mean level spacing $\delta$. 
By using the Kwant code, we numerically obtain  $E_T/\delta=11.1\pm0.5$ and $8.8\pm0.4$ 
for $W=4$ and 8, respectively \cite{kwant}.
% $E_T$ can be numerically obtained by using Kwant code \cite{thouless,kwant} whose values in the current 
% cases for $W=4$, 8 are $E_T/\delta=11.1\pm0.5$ and $8.8\pm0.4$, respectively.
There are three interesting regions. One is the so-called free diffusion limit: 
$N=\Delta E/\delta\gg E_T/\delta$ where $\Sigma_2$ is given by 
$\Sigma^D_2=(\sqrt{2}/(12\pi^3))(\delta/E_T)^{3/2}N^{3/2}$ \cite{book}, see  
the violet ($W=4$) and green ($W=8$) lines without any 
fitting parameter. Our numerical data agree well with the theory.
% and the level statistics of WSMs and DMs are the same in this (large $N$) region. 
In another limit of $\Delta E\ll \delta$, $\Sigma_2(\Delta E)$ 
follow the Poisson behavior $\Sigma_2(\Delta E)=N$ because the level repulsion will
not play any role when the probability of having two levels in $\Delta E$ is negligible. 
Our numerical data for both WSM and DM agree again with this expectation, evident by the 
fact that black line of $\Sigma_2(\Delta E)=N$ passes through all data points for $N<0.1$.
Thus, the level statistics of WSMs and DMs are the same in the two regions.
\par

The intermediate region is the so-called ergodic limit of $\delta<\Delta E<E_T$, 
where the level statistics of WSMs and DMs are completely different. 
For DMs, the level repulsion dominates the statistics, leading to the 
Wigner-Dyson statistics, $\Sigma_2=(\ln (2\pi N)+1+e^\gamma)/\pi^2$ 
(blue curve), where $\gamma\simeq 0.577$ is the Euler's constant \cite{mehta}.
Indeed, our data for $W=8$ agree with this prediction very well for $0.5<N<10$. 
Remarkably, a deviation from the Wigner-Dyson statistics is obviously seen for $W=4$, 
since the overlap of wavefunctions near WNs are much less than the normal extended 
states. The collective organization of wavefuctions near WNs is much weaker, and 
$\Sigma_2$ shows a linear behaviour in our data with a universal slope $\chi=0.2\pm0.1$
(red line) \cite{supp1}.
\par

% We would like to make a few remarks before the conclusion. 
\emph{Remarks.$-$}(1) A natural question is whether the physics and critical exponents 
such as $\gamma_0=0.62$ and $\nu=0.89$ are model-independent \cite{model}. As shown in 
the Supplemental Materials \cite{supp}, the answer is yes, at least within the models used.
% they are at least with the models that we worked on. 
Furthermore, $\gamma_0$ and $\nu$ independently obtained satisfy well the 
relationship of $\gamma_0=1-1/(\nu d)$ from the theory \cite{admirlin,altshulerb1}.
(2) The anomalous dimension $\Delta_2=D-d=-0.82$ agrees well with an analytical result  
\cite{syzranovsv}. (3) The estimated $\nu=0.89$ is consistent with result of the 
finite-size scaling of DOSs \cite{Q2,Q3} on double-WN models, but smaller than those by  
renormalization group calculations on single-WN models \cite{singleWN}.
%(2) Our estimated $\nu=0.89\pm0.05$ is smaller than renormalization group (RG) 
%calculations, $\nu=1$ for one-loop \cite{goswamip}, $\nu=1.14$ for two-loops 
%\cite{royb1}, and $\nu=1.47\pm0.03$ from numerical RG \cite{sbierskib1}.
%However, the calculation was based on the single-WN model in the $2+\epsilon$ dimension. 
%In reality, WNs must appear in pairs, and disorder will surely couple different WNs together \cite{suy}.
%Thus, the applicability of such a RG calculation is questionable. 
%In fact, our results are consistent with those ($\nu=0.86$) obtained from the double-WNs 
%model by the finite-size scaling of the density of states \cite{Q2,Q3}. 
(4) The spectral compressibility $\chi$ is related to the fractal dimension 
$D$ by $\chi=(d-D)/(2d)$ \cite{admirlin}. Numerical values of 
$\chi=0.2$ and $D=2.18$ agree with this relationship within numerical errors.
(5) Cold atom systm supporting WNs \cite{coldatom1,coldatom2} is the ideal platform 
to verify our theoretical prediction of the novel level statistics \cite{slzhu1}.
\par

In conclusion, the WNs ($E=0$) in weakly disordered WSMs are an isolated critical point 
in the sense that the zero energy wavefunction is a fractal of dimension $2.18$. 
There exists a correlation length diverging as $\xi=\xi_0|E|^{-\nu}$ with $\nu=0.89$ 
near the WNs. Wavefunctions exhibit fractal structures at the length scale smaller 
than $\xi$, and homogeneous strucutres at the length scale larger than $\xi$. 
Near the WNs and in a narrow energy window smaller than $[-\epsilon_c,\epsilon_c]$
($\epsilon_c=(\xi_0/L)^{-\nu}$), the nearest-neighbor level spacing distribution is 
well described by the critical level statistics of $P_c(s)=C_1s^2\exp[-C_2s^{2-\gamma_0}]$ 
with $\gamma_0=0.62\pm0.07$, in contrast to the Wigner-Dyson distribution far from the WNs. 
Similar conclusion is obtained for the level number variance $\Sigma_2(\Delta E)=\chi N$ 
around the WNs with a universal spectral compressibility $\chi=0.2\pm0.1$ .
The fractal nature and the level statistics of WNs thus provide authentic 
fingerprints of disordered WSMs.
\par

\begin{acknowledgments}

C.W. would like to thank Ying Su for valuable discussions of the phase
diagram of disordered WSMs.
This work is supported by the National Natural Science Foundation
of China (Grants No.~11374249 and 11704061) and Hong Kong RGC (Grants No.~16301518 and 
16300117). C.W. is supported by UESTC and the China Postdoctoral Science Foundation
(Grants No.~2017M610595 and 2017T100684). P.Y. is supported by the 
National Natural Science Foundation of China under Grant No.~11604041 and 
the National Thousand-Young-Talent Program of China. 

\end{acknowledgments}


\begin{thebibliography}{99}
\bibitem{symmetry} $\beta=1$ for systems with time-reversal symmetry and spin-rotation symmetry is 
called the Gaussian orthogonal ensemble; $\beta=2$ for systems without time-reversal symmetry 
is called the Gaussian unitary ensemble; $\beta=4$ for systems with time-reversal 
symmetry and no spin-rotation symmetry is called the Gaussian symplectic ensemble.
\bibitem{dos}The predicted vanishing DOSs at WNs have also 
attracted many numerical studies \cite{Q2,Q4,Q7,Q3,beras,pixleyjh2,Ziegler1}.
Most of those agreed with the zero DOS at WNs, but some recent works \cite{pixleyjh2,Ziegler1} 
concluded that it cannot exist at nonzero disorder due to rare region effects, and no 
WSM phase is allowed at an arbitrary weak disorder if zero DOS at WNs is demanded.
\bibitem{singleWN} We obtain $\nu=0.89\pm0.05$. The value is smaller than those from
renormalization group calculations, e.g., $\nu=1$ for one-loop \cite{Q1}, 
$\nu=1.14$ for two-loops \cite{royb}, and $\nu=1.47\pm0.03$ from numerical 
data \cite{sbierskib1}. Note that those calculations were based on the 
single-WN model in the $2+\epsilon$ dimension. In reality, WNs must appear 
in pairs. The white-noisy disorders will surely couple different WNs 
together \cite{suy}. Thus, the applicability of the renormalization group 
approach to finding critical exponent $\nu$ are questionable for the spatial 
uncorrelated disorders considered here.
\end{thebibliography}
\end{document}